\documentstyle[12pt]{article}

\def\thefootnote{\fnsymbol{footnote}}

\newcommand{\eq}{\begin{equation}}
\newcommand{\en}{\end{equation}}
\newcommand{\eqa}{\begin{eqnarray}}
\newcommand{\ena}{\end{eqnarray}}

\def\ee#1{{\rm e}^{#1}}
\def\trace{{\rm Tr}\hskip 1pt}

\def\ibes#1{{I_{#1}(\beta_t)\over I_1(\beta_t)}} 
\def\ibessp#1{{I_{#1}(\beta_s)\over I_1(\beta_s)}}

\newcommand{\NP}[1]{Nucl.\ Phys.\ {\bf #1}}
\newcommand{\PL}[1]{Phys.\ Lett.\ {\bf #1}}

\newcommand{\PR}[1]{Phys.\ Rev.\ {\bf #1}}
\newcommand{\PRL}[1]{Phys.\ Rev.\ Lett.\ {\bf #1}}

\begin{document}
\begin{titlepage}
\vskip0.5cm
\begin{flushright}
DFTT 57/95\\
September 1995
\end{flushright}
\vskip0.5cm
\begin{center}
{\Large\bf Recent results in high temperature
\footnote{Talk given at the Enrico Fermi International School
of Physics on "Selected Topics in Non-Perturbative QCD",
June 1995, Varenna, Italy.}}
\\
{\Large\bf  Lattice Gauge Theories}
\end{center}
\vskip 1.3cm
\centerline{M. Caselle}
\vskip 1.0cm
\centerline{\sl  Dipartimento di Fisica 
Teorica dell'Universit\`a di Torino}
\centerline{\sl Istituto Nazionale di Fisica Nucleare, Sezione di Torino}
\centerline{\sl via P.Giuria 1, I-10125 Torino, Italy
\footnote{e--mail: caselle~@to.infn.it}}
\vskip 1.cm

\begin{abstract}
\vskip0.2cm
We review some analytic results on the deconfinement transition in pure lattice
gauge theories. In particular we discuss the relationship between the
deconfinement transition in the $(d+1)$-dimensional $SU(2)$ model and the
magnetization transition in the $d$-dimensional Ising model. This analysis
leads to a precise estimate of the deconfinement temperature which agrees well
with that obtained with a Montecarlo simulation in the case in which the 
lattice has only one link in the compactified time direction.

\end{abstract}
\end{titlepage}

\setcounter{footnote}{0}
\def\thefootnote{\arabic{footnote}}

% 1.
\section{Introduction}
One of the most interesting predictions of QCD  is the existence of a 
deconfinement transition at some critical temperature $T_c$.
Finding a precise characterization of this transition is, however, still 
an open problem. In particular there are two main open questions: the first one
is the identification of the order of the transition and, in case it 
is of second order, of its critical indices. The second one is the precise
location of the deconfinement temperature.
The natural framework to pose these questions is that of the finite 
temperature Lattice Gauge Theories (LGT). In this framework, a seminal 
contribution was given more than ten years ago by B. Svetitsky and 
L.Yaffe~\cite{sy} in the case of {\sl pure} gauge theories. 
They showed that, if the deconfining transition of a 
given $(d+1)$ dimensional gauge theory is of 
second order, then its universality class should coincide with that of 
the $d$-dimensional spin model , with symmetry group the center of 
the original gauge group. This result is usually known as the 
``Svetitsky-Yaffe (SY) conjecture'' and has been confirmed in these last 
ten years by several Montecarlo simulations. It must be noticed however 
that the SY conjecture  gives no information on the location of the
deconfinement transition. Trying to answer to this last question will be the
main goal of the present contribution. In particular we shall see how far one
can go in trying to estimate the critical temperature by using only analytic
methods. We shall concentrate only on pure gauge theories with gauge group 
$SU(2)$, but most of our results can be straightforwardly extended to $SU(N)$
models with $N>2$.

During these last years the best estimates of the
critical temperature have been obtained by means of Montecarlo simulations,
which are certainly the most powerful tool to extract quantitative results from
LGT. However we think that it is important in itself
to have some independent analytic estimate of the
location of the critical point, besides the outputs of the computer
simulations, to reach a deeper theoretical understanding of the deconfinement
transition. 
The attempts to obtain analytically the critical
temperature have a rather long history, starting  more than ten years
ago~\cite{og,djk,gk,gw,ps}.
However the strategy has always been essentially the same: first construct an
effective action in terms of the Polyakov loops (which, as we shall see below,
 are the relevant
dynamical variables in the physics of deconfinement for pure gauge theories).
Second, use a mean field approximation to extract the critical coupling.
A common feature of all these attempts was that
the effective actions were always
 constructed neglecting the spacelike part of the
action. As a consequence it was impossible to reach 
a consistent continuum limit for the critical temperature. 
Moreover, as a consequence of the mean field approximation, the estimates of the
critical temperature were in general affected by large systematic errors.

The aim of the present contribution is to show that it is indeed possible
to overcome these two difficulties. First we shall construct in the $SU(2)$
case an improved effective action which takes also into account the spacelike
part of the original Wilson action. Second we shall avoid the mean field
approximation, and shall instead obtain the critical 
temperature by mapping (following the SY
conjecture) the gauge theory to a suitable Ising-like model, and then using the
fact that the critical temperature of the Ising model is known exactly in 
$d=2$, and with very high precision in $d=3$.
Let us stress that this is not the only possible way to avoid the mean field
approximation.
 Another interesting possibility
 is to study the $SU(N)$ models in  the 
 $N\to\infty$ limit where one can use some results recently obtained in
the context of random matrix models and two-dimensional exactly solvable gauge
theories to discuss the deconfinement transition and fix the critical
temperature. We shall not describe here this approach, the interested
reader can find in~\cite{bcdmp} a comprehensive discussion on the 
subject.

This contribution is organized as follows. Sect.2  will be devoted to 
a brief introduction to finite temperature LGT and to the Svetitsky-
Yaffe conjecture. In sect.3 we shall
 construct the  effective action, 
which we shall then use in sect.4
to extract the critical deconfinement temperature.
Finally sect.5 will be devoted to some concluding remarks.

%2

\section{Finite Temperature LGT}

\subsection{General setting}
Let us consider a pure gauge theory with gauge group $SU(N)$, defined
on a $d+1$ dimensional cubic lattice.
In order to describe a finite temperature LGT,
we have to impose periodic
boundary conditions in one direction (which we shall call from now
on ``time-like'' direction), while the boundary conditions in the
other $d$ direction (which we shall call ``space-like'') can be
chosen freely.
We take a lattice of $N_t$ ($N_s$) spacings in the time (space)
direction, and we work with the pure gauge theory, containing
only gauge fields described by the link
variables $U_{n;i} \in SU(N)$, where $ n \equiv (\vec x,t)$
denotes the space-time position of the link and $i$ its direction.
It is useful to choose different bare couplings in the time and space
directions. Let us call them $\beta_t$ and $\beta_s$ respectively.
The Wilson action is then
\eq
S_W=\sum_{n}~\frac{1}{N}{\rm Re}\left\{\beta_t\sum_i~{\rm Tr_f}(U_{n;0i})
+\beta_s\sum_{i<j}~{\rm Tr_f}(U_{n;ij})\right\}~~~,
\label{wilson}
\en
where ${\rm Tr_f}$ denotes the trace in the fundamental representation
and $U_{n;0i}$ ($U_{n;ij}$) are the time-like (space-like)
plaquette variables, defined as usual by
\eq
U_{n;ij}=U_{n;i}U_{n + i;j}
U^\dagger_{n + j;i}U^\dagger_{n;j}~~~.
\en

In the following we shall call $S_s$ ($S_t$) the space-like (time-like)
part of $S_W$. $\beta_s$ and $\beta_t$ are related to the (bare) gauge
coupling $g$ and to the temperature $T$ by the usual relations

\eq
\frac{4}{g^2}=a^{3-d}\sqrt{\beta_s\beta_t}~~~,
\hskip 1cm
T=\frac{1}{N_ta}\sqrt{\frac{\beta_t}{\beta_s}}~~~,
\label{couplings}
\en
where $a$ is the space-like lattice spacing, while $\frac{1}{N_tT}$ is
the time-like spacing.

In a finite temperature discretization it is possible to define
gauge invariant observables which are topologically non-trivial,
as a consequence of the periodic boundary conditions in the
time directions.
The simplest choice is the Polyakov loop,
defined in terms of link
variables as
\eq
P(\vec x)\equiv
{\rm Tr_f}~P_{\vec x}= {\rm Tr_f} \prod_{t=1}^{N_t}(U_{\vec x,t;0})~~~.
\label{polya}
\en
In the following we shall call $P_{\vec x}$, ``open Polyakov line''.

As it is well known, the finite temperature theory has a new
global symmetry (unrelated to the gauge symmetry), with symmetry group
the center $C$ of the gauge group (in our case $Z_2$). The Polyakov loop
is a natural order parameter for this symmetry.

In $d>1$, finite temperature gauge theories admit a
deconfinement transition at $T=T_c$, separating the
high temperature, deconfined, phase ($T>T_c$) from the low
temperature, confining domain ($T<T_c$).
The high
temperature regime is characterized by the  breaking of the global
symmetry with respect to the center of the group. In this phase the
Polyakov loop has a non-zero expectation value, and it is
an element of the center of the gauge group. In the low temperature phase the
center symmetry is conserved and the expectation value of the Polyakov loop is
zero. The relevant feature of the Polyakov loop is that at the same time it is
also the order parameter for the deconfinement transition. In fact
its expectation value is related
to the free energy of an isolated, static quark as follows: 
\eq
<P>~~\propto~~~ \exp(-F_q)~~~~~.
\en
As a consequence, in the low temperature phase it would
require an infinite energy to create from the vacuum an isolated quark.
Hence in this phase quarks are confined. On the contrary, in the high 
temperature
phase isolated quark can exist: this is the deconfined phase.
The critical point in which the center symmetry  is broken can thus be
interpreted as the deconfinement transition. 
The corresponding critical temperature $T_c$ will be denoted in the following 
as the deconfinement temperature.

\subsection{Svetitsky-Yaffe conjecture}

The idea on which the SY conjecture is based is that, if one would be 
able to integrate out all the 
gauge degrees of freedom of the original $(d+1)$--dimensional
 model {\sl except those related to the Polyakov loops}
then the resulting 
effective theory for the Polyakov loops would be a $d$-dimensional spin 
system with symmetry group $C$. The deconfinement transition of the 
original model would become the order--disorder transition of the 
effective spin system.
This effective theory would obviously have very complicated
interactions, but Svetitsky and Yaffe were able to argue that all these 
interactions should be short ranged. As a consequence, if the 
transition point of the
effective spin system is of second order, near this critical point, 
where the correlation length becomes infinite, these short ranged 
interactions can be neglected, and the universality class of the 
deconfinement transition should coincide with that of the simple spin 
model with only nearest neighbour 
interactions and the same global symmetry group. 
As a consequence all the critical indices describing the two 
transitions  and all the adimensional ratios of 
scaling quantities should 
coincide in the limit. 
In particular the $(d+1)$-dimensional $SU(2)$ 
gauge theory, which is known to
have a second 
order deconfinement transition, is characterized by the same 
renormalization group fixed point of the $d$--dimensional Ising model.

Unfortunately the SY argument alone
 cannot help to fix the critical temperature, 
since the precise mapping between the Ising coupling and that of the 
original gauge model, requires taking into account exactly those short 
ranged interaction which we neglected above.

\subsection{The $SU(2)$ case: character expansion}

In the following we shall  concentrate on the $SU(2)$ model. There
are two important features which greatly simplify the analysis in this case.
The first one is that according to the $SY$ conjecture the
model can be mapped, at the deconfinement point into the spin Ising model, which
is exactly solved in $d=2$ and very well known in $d=3$. The second important
feature is that in the $SU(2)$ case the
character expansion (which plays an important role in the construction of the
effective action) is very easy to handle.

Let us briefly summarize few results.
The character of the group element $U$ in the $j^{th}$ representation 
is:
\eq
\chi_j(U)\equiv {\rm Tr_j}(U)=\frac{\sin((2j+1)\theta)}{\sin(\theta)}
\label{c1}
\en
where ${\rm Tr_j}$ denotes the trace in the $j^{th}$ representation and 
$\theta$ is defined according to the following parametrization of $U$ in 
the fundamental representation:
\eq
U=\cos(\theta){\bf 1} + i\vec\sigma\vec n \sin(\theta)
\label{deftheta}
\en
where $\vec n$ is a tridimensional unit vector and $\sigma_i$ are the 
three Pauli matrices. Notice, as a side remark, that with this 
parametrization the Haar measure has the following form:
\eq
DU=sin^2(\theta)\frac{d\theta d^2\vec n}{4\pi^2}
\en
and the Polyakov loop becomes $P(\vec x)=2\cos(\theta_{\vec x})$

The following orthogonality relations between characters hold:
\eq
\int D~U \chi_r(U)~\chi_s(U)=\delta_{r,s}
\label{c2}
\en

\eq
\sum_{r} d_r \chi_r(U~V^{-1})= \delta(U,V)
\label{c3}
\en

where $d_r$ denotes the dimensions of the $r^{th}$ representation:
$d_r=2r+1$.
In the following we shall use two important properties of the 
characters:

\eq
\int D~U \chi_r(U)~\chi_s(U^{-1}V)
=\delta_{r,s}\frac{\chi_r(V)}{d_r}~~,
\label{c4}
\en
\eq
\int D~U \chi_r(UV_1U^{-1}V_2)=\frac{1}{d_r}\chi_r(V_1)\chi_r(V_2)~~~.
\label{c5}
\en

The character expansion of the Wilson action has a particularly simple 
form:
\eq
e^{\frac{\beta}{2}{\rm Tr_f}(U)}=\sum_j 2(2j+1)\frac{I_{2j+1}(\beta)}
{\beta}\chi_j(U),~~~j=0,\frac{1}{2},1\cdots
\en
where $I_n(\beta)$ is the $n^{th}$ modified Bessel function.

In the following we shall often use the  normalized version of the 
character expansion in which
 the coefficient of the trivial representation is set to 1.
\eq
e^{\frac{\beta}{2}{\rm Tr_f}(U)}=G(\beta)\sum_j (2j+1)\frac{I_{2j+1}(\beta)}
{I_{1}(\beta)}\chi_j(U),~~~j=0,\frac{1}{2},1\cdots
\label{use1}
\en
where $G(\beta)=2I_1(\beta)/\beta$ is an irrelevant constant that we 
shall often omit.

%3

\section{Construction of the Effective Action}

In this section our goal is to construct an effective action for the 
finite temperature LGT in terms of the Polyakov loops only. This implies that
we must be able  to integrate out exactly all the 
spacelike variables so that the only remaining degrees of 
freedom at the end are exactly the Polyakov loops. Notice that in this 
way the resulting effective action would live in $d$ dimensions 
(one dimension less than the starting model). This is exactly along the 
line of the original Svetitsky-Yaffe program. In trying to follow such a 
program one must necessarily make some approximation.
In order to obtain a good approximation of the
original Wilson action, one must  identify
the physically relevant degrees of freedom, and then try to
  keep them unchanged when constructing
the effective action.
Following~\cite{sy} and~\cite{bcdmp}, we assume that the 
physics of the deconfinement transition is  dominated by the 
timelike plaquettes, and try to keep as far as possible unchanged this part of
the original action. Accordingly 
we treat the spacelike part of the Wilson action: $S_s$
 as a perturbation of the timelike part $S_t$ and take care of the 
contributions coming from $S_s$ by making a strong 
coupling expansion in $\beta_s$. The main difference with respect to the 
usual approach is that in this case the time-like part of action is 
treated exactly or, equivalently, that the expansion in $\beta_t$ is 
summed up to all orders. The only remaining expansion parameter is 
thus $\beta_s$.
 In particular the zeroth order in $\beta_s$
 will contain the timelike plaquettes only. It is not at all obvious 
that the integration over the spacelike links could be done to all 
orders in $\beta_t$, but it turns out that it can be done exactly in the 
framework of the characters expansion  {\sl order by order 
in $\beta_s$}. 
In particular we shall discuss the zeroth and first order in $\beta_s$ 
only, 
which will be enough to our purposes, but there is in principle no 
obstruction to go to higher orders. The result for any given order in 
$\beta_s$ can be expressed as an infinite sum over characters. 

Remarkably enough in the $N_t=1$ case this series can be summed exactly 
and the result can be written in a closed form. This is essentially due 
to the fact the if $N_t=1$ this same effective action can be obtained 
in a completely different way, using techniques typical of matrix models
(see below), thus allowing a non trivial  check of all our strong 
coupling results. Another interesting feature of the $N_t=1$ limit is that in
this case very precise Montecarlo estimates of $T_c$ exist, with which we can
compare our analytic  predictions. For instance, in
 (3+1) dimensions the critical coupling $\beta_c$ 
at which deconfinement occurs is estimated to be: 
 (for the $SU(2)$ model with $N_t=1$)
$\beta_c=0.8730(2)$~\cite{bems}. 
Such an impressive precision is due to that fact that in
the $N_t=1$ case (and {\sl only} in this case) one can simulate the gauge model
by using a cluster non-local algorithm (see~\cite{bems} for the details). 
This makes
the $SU(2)$, 
$N_t=1$ model a perfect laboratory to test our techniques, and we
shall concentrate on this case in sect.4 . We shall briefly
comment on the extension of our results to $N_t>1$ in sect.5 .

\subsection{Expansion in $\beta_s$ of the effective action}

The effective action $S_{\rm eff}$ for the Polyakov loops $P_{\vec x}\equiv
\prod_{t=1}^{N_t} V_{\vec x}$ is obtained integrating over all the 
spacelike degrees of freedom in the action (\ref{wilson}). As explained
previously, our approach is to consider the contributions from the spacelike
plaquettes up to a certain order in $\beta_s$ only. 
So, for our purposes, it will be convenient to expand separately the spacelike
and the timelike part of the action (\ref{wilson}):
\eqa
\label{gen1}
\ee{S_{\rm eff}} & = & \int\prod_{\vec x,t;i} DU_{\vec x,t;i}
 \exp S_W
  \nonumber \\
& = & \int\prod_{\vec x,t;i} DU_{\vec x,t;i} \hskip 2pt 
\prod_{\vec x, t ; i} \left(1 + \sum_{j={1\over2}}^\infty d_j
\ibes{2j+1} \chi_j(U_{\vec x, t;0i}) \right)\nonumber\\
& & \times \prod_{\vec x,t; i,j} \left(1 + \sum_{l={1\over2}}^\infty d_l
\ibessp{2l+1} \chi_l(U_{\vec x,t; ij})\right)~~.
\ena
Specifically, we work out here the effective action
up to $O(\beta_s^2)$. This means that in eq. (\ref{gen1}) we must look only at 
the terms containing at most a single space-like plaquette in the adjoint
representation, $\chi_1(U_{\vec x,t;ij})$, or two space-like plaquettes in the
fundamental, $\chi_{1\over 2}(U_{\vec x,t_1;ij}) \chi_{1\over 2}
(U_{\vec y,t_2;kl})$. 
Due to the orthogonality relations for characters, it easy to convince 
oneself that a pair of plaquettes in the fundamental representation 
do actually contribute to the integral only if they appear in the 
same spatial position (at two different times $t_1$ and $t_2$);
for the same reason a single fundamental plaquette cannot contribute.
We are thus lead to the following expression: 
\eqa
\label{gen2}
&&\exp(S_{\rm eff}) = \int\prod_{\vec x,t;i} DU_{\vec x,t;i}\hskip 2pt 
\prod_{\vec x,t; i} \left(1 + \sum_{j={1\over2}}^\infty d_j
\ibes{2j+1} \chi_j(U_{\vec x, t; 0i} ) \right)\nonumber \\
&&\times \left( 1 + \sum_{\vec x, i < j} \left[\sum_{t=1}^{N_t} 3 \ibessp{3}
\chi_1(U_{\vec x, t; ij}) + \sum_{t_1 < t_2} 
4 \left(\ibessp{2}\right)^2 
\chi_{1\over 2}(U_{\vec x, t_1; ij}) 
\chi_{1\over 2}(U_{\vec x, t_2; ij})\right]\right).
%\nonumber\\ 
\ena

In the following we shall calculate these integrals for a generic value of 
$N_t$

\subsection{Zeroth order approximation}

Let us first study the contribution which gives the 
$O(\beta_s^0)$ result and corresponds to the ``$1$'' in the second
factor of eq.(\ref{gen2}).
In this case the integral only contains
 the timelike part of the Wilson action:
\eq
\label{zero1}
\exp(S_0) = \int\prod_{\vec x,i;t}\hskip 2pt\left[DU_{\vec x,i;y}\hskip 2pt
\left(1+\sum_{j={1\over 2}}^\infty d_j\ibes{2j+1} \chi_j(U_{\vec x,i;t}
V_{\vec x + i;t} U_{\vec x,t+1;i}^\dagger V_{\vec x;t}^\dagger)
\right)\right], 
\en
and it is easy to integrate all the spacelike 
links. The reason is that each spacelike link only belongs to two 
timelike plaquettes, hence by making a character expansion, it can be 
exactly integrated out. Let us do this integration in two steps, for 
future commodity. First let us integrate (by using eq.(\ref{c4}))
all the spacelike links except 
the lowermost ones (which, due to the periodic boundary conditions 
coincides with the uppermost). We obtain:
\eq
\exp(S_0) =  \prod_{\vec x,i}\left(1+\sum_{j={1\over 2}}^\infty
\left[\ibes{2j+1}\right]^{N_t}
~\chi_j\left(U_{\vec x;i}P_{\vec{x}+i}
U^\dagger_{\vec x;i}P^\dagger_{\vec x}\right)\right)~~.
\label{a1}
\en
where $P_{\vec x}$ is the open Polyakov line (whose trace is the 
Polyakov loop) in the site $\vec x$ and $U_{\vec x;i}$ are the remaining
lowermost  
spacelike links. Integrating also on $U_{\vec x;i}$ 
(this time, by using eq.(\ref{c4})) we end up with
\eq
\label{a2}
\exp(S_0) = \prod_{\vec x,i}\left(1+\sum_{j={1\over 2}}^\infty
\left[\ibes{2j+1}\right]^{N_t}
~\chi_j(P_{\vec{x}+i}) ~\chi_j (P^\dagger_{\vec{x}}) 
\right)~~.
\en
Let us define, for future convenience, the link\footnote{The links we are 
referring to are those of the $d$-dimensional spatial lattice, corresponding
to one ``horizontal'' slice in the original $d+1$-dimensional 
lattice} element of $\exp(S_0)$ as follows:
\eq
C^0_{\vec x; i} \equiv \sum_{j=0}^\infty
\left[\frac{I_{2j+1}(\beta_t)}{I_1(\beta_t)}\right]^{N_t}
~\chi_j (P_{\vec x+i})\chi_j (P^\dagger_{\vec{x}})~~.
\label{n2}
\en
It is now evident that this basic element, which will be denoted also as 
$C^0_{\vec x, i} =
C^0(\theta_{\vec x},\theta_{\vec x+ i})$, depends only on  
$\theta_{\vec x}, \theta_{\vec x +i}$, which are  the invariant angles 
for the Polyakov lines 
$P_{\vec x},P_{\vec x +i}$ in the sites joined by the link. 
Indeed from now on we will always 
assume to have gauge-rotated the Polyakov lines to be diagonal:
\eq
\label{zero2}
P_{\vec x}=\left(\begin{array}{cc} e^{i\theta_{\vec x}} & 0 \\
 0 & e^{-i\theta_{\vec x}}  \end{array}\right)~~~.
\en
With these definitions the zero$^{\rm th}$-order action (\ref{a2})
is simply given by
\eq
\label{zero3}
\exp(S_0) = \prod_{\vec x ; i} C^0_{\vec x;i}~~~~.
\en

However let us stress that the action that was generally used in
the previous attempts to obtain mean field estimates of the deconfinement
temperature, was actually a simplified version 
(truncated at the first representation) of eq.(\ref{a2}):
\eq
S_{p}(\beta_t)=\sum_{\vec x}\left\{\beta_t\sum_{i=1}^d
~\cos(\theta_{\vec x})~\cos(\theta_{\vec{x}+i})\right\}~~~.
\label{seff}
\en

\vskip 0.3cm
It is interesting to notice that in the
 $N_t=1$ case the character expansion contained in eq.(\ref{a2})
can be summed exactly. This can be easily seen by writing the explicit 
form for the characters in eq.(\ref{a2}) :
\eq
\exp(S_0) =  \prod_{\vec x,i}\left(1+\sum_{j={1\over 2}}^\infty
\left[\frac{I_{2j+1}(\beta_t)}{I_1(\beta_t)}\right]
~\frac{\sin[(2j+1)\theta_{\vec x}]~\sin[(2j+1)\theta_{\vec x+i}]}
{\sin(\theta_{\vec x})~\sin(\theta_{\vec x+i})}\right)~~.
\label{a2bis}
\en

Then using the relation:
\eq
2~\sin[(2r+1)\theta_{\vec x}]~\sin[(2r+1)\theta_{\vec x+i}]=
\cos[(2r+1)(\theta_{\vec x}-\theta_{\vec x+i})]-
\cos[(2r+1)(\theta_{\vec x}+\theta_{\vec x+i})]
\label{a2ter}
\en
and the well known expansion:
\eq
e^{\beta\cos{\theta}}=I_0(\beta)+2\sum_{k=1}^\infty 
I_k(\beta)\cos(k\theta)~~~,
\label{a2quat}
\en
it is easy to obtain:
\eq
\exp(S_0) =  \prod_{\vec x,i}
\frac{e^{\beta_t \cos(\theta_{\vec x}-\theta_{\vec x+i})}-
e^{\beta_t \cos(\theta_{\vec x}+\theta_{\vec x+i})}}
{4I_1(\beta_t) \sin(\theta_{\vec x}) \sin(\theta_{\vec x+i})}~~.
\label{a3}
\en

\subsection{First order approximation}

The $O(\beta_s^0)$ effective action (\ref{zero3}) 
contains just nearest-neighbour
interactions between the Polyakov loops. We shall show below that
 $O(\beta_s^2)$ contributions  to the effective action are of plaquette type,
 namely they involve all the four invariant angles of the Polyakov lines 
which belong to a given plaquette. This interaction is more general than the
nearest-neighbour one, but it is still short ranged, in agreement with the
hypothesis which is at the basis of the SY conjecture discussed above.

\subsubsection{The adjoint representation term.}

To calculate the contribution coming
 from the adjoint representation term, we have to
select in eq.(\ref{gen2}) the term:

\eqa
 \phantom{p} &
3 \ibessp{3}
\int \prod_{\vec x,t;i} DU_{\vec x,t;i}\hskip 2pt 
\prod_{\vec x, t; i} \left(1 + \sum_{j={1\over2}}^\infty d_j
\ibes{2j+1} \chi_j(U_{\vec x, t; 0i} ) \right) \times \nonumber \\
\times & 
\sum_{\vec x, i < j}
\sum_{t=1}^{N_t}
\chi_1(U_{\vec x, t; ij})~~~.
\label{adj1}
\ena
To study the integral (\ref{adj1}), we first  note that all the spacelike
plaquettes in the same spatial position give evidently the same contribution,
regardless of the time $t$; therefore the sum over the time positions in
(\ref{adj1}) simply results in a $N_t$ factor. Secondly, it is convenient to
use the following relation for the $SU(2)$ characters:
\eq
\label{chi1}
\chi_1 = (\chi_{1\over 2})^2 - 1~~~ .
\en
The ``$-1$'' simply reproduces the zeroth order 
term, and gives a renormalization of order $\beta_s^2$ to such 
contribution.
The integral along the plaquette can now be decoupled into integrals over a
single link matrix, by writing explicitly 
$[\chi_{1 \over 2}(U_{\vec x,t ; ij})]^2$ as a product of
elements (in the fundamental representation) of the link matrices.
Thus eq. (\ref{adj1}) can be rewritten in terms of the following integrals over
the unitary spacelike link matrix $U$:
\eq
\label{beq}
B_{\alpha\beta\gamma\delta}(P_{\vec x},P_{\vec x+ i}) = \int DU \hskip 3pt 
\left(1+\sum_{j={1\over 2}}^\infty d_j \left[\ibes{2j+1}\right]^{N_t}
\chi_j(U P_{\vec x +i} U^\dagger P_{\vec x}^\dagger )\right) U_{\alpha\beta} 
U^\dagger_{\gamma\delta}
\en
where $\alpha,\ldots=1,2$ are the indices of the $U$ matrix in the 
fundamental representation.
Making use of the invariance of the measure and of the argument of $\chi_j$ 
in eq. (\ref{beq}) under the transformations 
\begin{equation}
U_{\alpha\beta} \rightarrow \omega_{\alpha\alpha}~U_{\alpha\beta}~,
\phantom{ppppp}
U^\dagger_{\gamma\delta} \rightarrow  
U^\dagger_{\gamma\delta}~ (\omega^{-1})_{\delta\delta}
\label{omegasym1}
\end{equation}
and
\begin{equation}
U_{\alpha\beta} \rightarrow U_{\alpha\beta} ~\omega_{\beta\beta}~,
\phantom{ppppp}
U^\dagger_{\gamma\delta} \rightarrow (\omega^{-1})_{\gamma\gamma}~
U^\dagger_{\gamma\delta}
\label{omegasym2}
\end{equation}
(where the diagonal unitary matrix $\omega$ has the property $\omega^2 = 1$),
one can conclude that 
$B_{\alpha\beta\gamma\delta} = 0 $ unless $\beta = \gamma$ and $\delta =
\alpha$. As a consequence, the integral (\ref{beq}) depends on the invariant
angles of the Polyakov loop only and can be written as follows:
\eq
\label{beq2}
B_{\alpha\beta\gamma\delta}(P_{\vec x}, P_{\vec x + i})
\equiv B_{\alpha\beta\gamma\delta}(\theta_{\vec x}, \theta_{\vec x + i})
=\delta_{\beta\gamma}\delta_{\delta\alpha}
~C_{\alpha\beta}(\theta_{\vec x}, \theta_{\vec x + i})
\en 
(no summation over repeated indices).
Moreover, it is not difficult to show that $C_{\alpha\beta}$ is a real
symmetric matrix.
By using these results, we can write the contribution (\ref{adj1}) to the
effective action only in terms of the invariant angles of the Polyakov line
$P_{\vec x}$. Eq. (\ref{adj1}) becomes:
\eq
3 N_t \frac{I_3(\beta_s)}{I_1(\beta_s)} ~
\biggl[ \prod_{\vec x, i} C^0_{\vec x,i} \biggr] ~
\biggl[
\trace [\widehat{C}(\theta_{\vec x},\theta_{\vec x +i})
\widehat{C}(\theta_{\vec x +i},\theta_{\vec x + i+j})
\widehat{C}(\theta_{\vec x +i+j},\theta_{\vec x + j})
\widehat{C}(\theta_{\vec x + j},\theta_{\vec x})] - 1
\biggr]
\label{adj2}
\en
The subtraction of the term (-1) in (\ref{adj2}) is due to the term (-1) in
(\ref{chi1}), whereas the matrices 
$\widehat{C}(\theta_{\vec x},\theta{\vec x + i})$ 
are the normalized version of $C$:
\eq
\label{chatadj}
\widehat{C}_{\alpha,\beta}(\theta_{\vec x},\theta_{\vec x+ i}) = 
{C_{\alpha,\beta}(\theta_{\vec x},\theta_{\vec x+ i})
\over C^0_{\vec x; i}}
\en

The last step is the explicit evaluation of the matrix elements 
$C_{\alpha\beta}$. This calculation is described in the Appendix. The final
result is:

\eqa
\label{trigo1}
C_{11}=C_{22} & = & \frac{1}{2}(C_0+C_1)\nonumber\\
C_{12}=C_{21} & = & \frac{1}{2}(C_0-C_1)
\ena
with:
\eqa
C_1 & = & {1\over {2\sin\theta_{\vec x}\sin\theta_{\vec x +i}}}
\biggl\{\sum_{j=0}^\infty 
\biggl[\chi_j(\theta_{\vec x})\chi_j(\theta_{\vec x+i})
\frac{I_{2j+2}(\beta_t)^{N_t}-I_{2j}(\beta_t)^{N_t}}
{(2j+1)I_1(\beta_t)^{N_t}}\nonumber\\
& & +2\cos\left[(2j+1)\theta_{\vec x}\right]\cos\left[(2j+1)\theta_{\vec x +i}
\right]
\left(\frac{I_{2j+1}(\beta_t)}{I_1(\beta_t)}\right)^{N_t}\biggr]+
\left[\frac{I_0(\beta_t)}{I_1(\beta_t)}\right]^{N_t}\biggr\}
\label{ab19}
\ena
$C_0$ was defined in eq.(\ref{n2}), and, as expected,
$C_{11}+C_{12}=C_0$.

Eq.(\ref{ab19}) could seem a bit complicated, 
but it is actually very 
easy to  implement it in a mean field analysis or in the SY type
 mapping described in
the next section.

\vskip 0.3cm
In the $N_t=1$ case
the sum over the representations can be performed exactly, and a closed 
expression for the $C_{\alpha\beta}$ coefficients can be obtained.
This can be done by using the identity:
\eq
\label{idbes1}
I(\beta)_{n-1}-I(\beta)_{n+1}=2nI(\beta)_{n}
\en
and eq(\ref{a2quat}).
The result is:
\eqa
C_{11}(N_t=1) & = &   
\frac{e^{\beta_t \cos(\theta_{\vec x} -\theta_{\vec x +i})}}
{4I_1(\beta_t) \sin(\theta_{\vec x}) \sin(\theta_{\vec x +i})}
-\frac{e^{\beta_t \cos(\theta_{\vec x} -\theta_{\vec x +i})}-
e^{\beta_t \cos(\theta_{\vec x} +\theta_{\vec x +i})}}
{8\beta_t I_1(\beta_t) \sin^2(\theta_{\vec x}) 
\sin^2(\theta_{\vec x +i})}\nonumber\\
%\label{an1}
C_{12}(N_t=1) & = &  
\frac{e^{\beta_t \cos(\theta_{\vec x} -\theta_{\vec x +i} )}-
e^{\beta_t \cos(\theta_{\vec x} +\theta_{\vec x +i})}}
{8\beta_t I_1(\beta_t) \sin^2(\theta_{\vec x}) 
\sin^2(\theta_{\vec x +i})}-
\frac{e^{\beta_t \cos(\theta_{\vec x}+\theta_{\vec x +i})}}
{4I_1(\beta_t) \sin(\theta_{\vec x}) \sin(\theta_{\vec x +i})}~~.
\label{an2}
\ena

\subsubsection{Pair of fundamental representations.}

To calculate the contribution coming from a pair of fundamental 
representations, we have to
select the last term in  (\ref{gen2}) :

\eqa
 \phantom{p} &
4 \left(\ibessp{2}\right)^2
\int \prod_{\vec x,t;i} DU_{\vec x,t;i}\hskip 2pt 
\prod_{\vec x, t; i} \left(1 + \sum_{j={1\over2}}^\infty d_j
\ibes{2j+1} \chi_j(U_{\vec x, t; 0i} ) \right) \times \nonumber \\
\times & 
\sum_{\vec x, i < j}
\sum_{t_1<t_2}
\chi_{\frac{1}{2}}(U_{\vec x, t_1; ij})
\chi_{\frac{1}{2}}(U_{\vec x, t_2; ij})~~~.
\label{fund2}
\ena

This term can be evaluated following the same pattern of the 
calculation in the adjoint representation case. 
It is however important to notice that in the $N_t=1$ limit this
contribution {\sl exactly disappears}.
Since this is the limit in which we shall be
interested in the following  we shall omit here the details of this
calculation which can be found in~\cite{bcdp}. 

\subsection{Independent approach to $N_t=1$} 
The interesting feature of the $N_t=1$ limit is that in this case the 
theory that we are studying becomes a particular instance of the  
Migdal-Kazakov model. This connection was already noticed in~\cite{cdp}
and was the origin of our previous analysis in the $N\to\infty$ 
limit~\cite{bcdmp}.
 All the integrals that we have described in the 
previous sections can be directly evaluated in this case as particular 
instances of a nontrivial generalization of the so called Itzykson-Zuber 
integral,
evaluated in~\cite{kmsw}. 
We refer the reader to~\cite{kmsw} for a comprehensive discussion on this
interesting subject 
and simply report here the results which are useful for our analysis:

zeroth order contribution:
\begin{eqnarray}
\int d U_{\vec{x};i} \exp \left\{ {\beta_t \over 2}
{\rm Tr}_f \left( V(\vec{x}) U_{\vec{x};i} V^\dagger(\vec{x}+i)
U^\dagger_{\vec{x};i} \right) \right\} & = &
\nonumber \\
= { e^{\beta_t \cos(\theta_{\vec{x}} -\theta_{\vec{x}+i})}
- e^{ \beta_t \cos(\theta_{\vec{x}} +\theta_{\vec{x}+i})}
\over 2 \beta_t \sin(\theta_{\vec{x}}) \sin(\theta_{\vec{x}+i}) }
& \phantom{=}~. &
\label{a6}
\end{eqnarray}

The first order contributions can be extracted by the correlators 
defined in~\cite{kmsw} as:
\begin{eqnarray} 
& &
\left<
\left( U_{\vec{x};i} \right)_{\mu,\nu} \left( U^\dagger_{\vec{x};i}
\right)^{\rho,\sigma} \right>
 \nonumber \\
&\equiv  &
\frac{ \int d U_{\vec{x};i} \exp \left\{
{\beta_t \over 2} {\rm Tr}_f \left( V(\vec{x}) U_{\vec{x};i} 
V^\dagger(\vec{x}+i)
U^\dagger_{\vec{x};i} \right) \right\} 
\left( U_{\vec{x};i} \right)_{\mu,\nu} \left( U^\dagger_{\vec{x};i}
\right)^{\rho,\sigma} }
{\int d U_{\vec{x};i} \exp \left\{
{\beta_t \over 2} {\rm Tr}_f \left( V(\vec{x}) U_{\vec{x};i} 
V^\dagger(\vec{x}+i)
U^\dagger_{\vec{x};i} \right) \right\} }~.
\label{a7}
\end{eqnarray}
It is easy to see that these correlators must be of diagonal form, 
namely:
\eq
\left<
\left( U_{\vec{x};i} \right)_{\mu,\nu} \left( U^\dagger_{\vec{x};i}
\right)^{\rho,\sigma} \right>
 =  \delta_\mu^\sigma \delta_\nu^\rho \hat C_{\mu,\nu} (\vec{x};i)
\en
where the $\hat C_{\mu,\nu}(\vec{x};i)$ are equivalent, apart from the 
different normalization, to our $C_{kl}$ matrix elements. They turn out 
to be~\cite{kmsw}:

\begin{eqnarray}
\hat C_{1,1}(\vec{x};i) & = & \hat C_{2,2}(\vec{x};i) = 
{ 2\beta_t \sin(\theta_{\vec{x}}) \sin(\theta_{\vec{x}+i}) - 
 \left( 1 - e^{-2\beta_t \sin(\theta_{\vec{x}}) \sin(\theta_{\vec{x}+i})}
\right) \over
\left( 1 - e^{-2\beta_t \sin(\theta_{\vec{x}}) \sin(\theta_{\vec{x}+i})}
\right) \left( 2\beta_t \sin(\theta_{\vec{x}}) \sin(\theta_{\vec{x}+i})
\right) }   
\nonumber \\
\hat C_{1,2}(\vec{x};i) & = &\hat C_{2,1}(\vec{x};i)  = 
{1 - e^{-2\beta_t \sin(\theta_{\vec{x}}) \sin(\theta_{\vec{x}+i})} 
\left(1+ 2\beta_t \sin(\theta_{\vec{x}}) \sin(\theta_{\vec{x}+i})
\right)
\over
\left( 1 - e^{-2\beta_t \sin(\theta_{\vec{x}}) \sin(\theta_{\vec{x}+i})}
\right) \left( 2\beta_t \sin(\theta_{\vec{x}}) \sin(\theta_{\vec{x}+i})
\right) }~~.
\label{a8}
\end{eqnarray}

It is now only matter of straightforward algebra (one must also take into
account eq.(\ref{chatadj}) and the normalization constant $G(\beta)$ defined 
in (\ref{use1}) ) to show that these 
expressions eq.(\ref{a6}) and (\ref{a8}) are exactly equivalent to our 
results (\ref{a3})  (\ref{an2}).

%4

\section{Determination of $T_c$}
As we discussed in the introduction, the standard approach to the determination
of the critical temperature would be at this point
 a mean field analysis of the above
constructed effective action. However this approach is rather
unsatisfactory. For instance if we take the standard mean field approximation 
of the zeroth order action, truncated at the first representation, 
eq.(\ref{seff})
 ( for which
the calculation can be performed exactly, see for instance~\cite{gk})
 it is easy to see that the resulting critical
coupling in the $N_t=1$ case is $\beta_c=2/d$. In  (3+1) dimensions we know that
$\beta_c=0.8730(2)$~\cite{bems},
and it is clear that the standard mean field
derivation which gives in this case $\beta_c=0.666...$ 
is largely unsatisfactory.
It is possible to improve this result keeping the full effective action instead
of its truncated version, and improving the mean field approximation. This gives
a much better result, which however always remains between 5 and 10 \% below the
Montecarlo result (see~\cite{bcdp} for a discussion of this approach). 

In this section we want to discuss a completely 
different approach, which makes explicit use of the mapping between the 
$SU(2)$
model and the $d$-dimensional Ising model and turns out to be much more 
powerful of the mean field approach. 
 Up to our knowledge the 
only attempt along this line was made by J. Polonyi and K.Szlachanyi 
in~\cite{ps}, but since they were constrained to keep in the various 
stages of their analysis only the very first order in $\beta_t$ (and to 
neglect $\beta_s$) their result was rather unsatisfactory. We shall review 
their approach below. 
The novelty of our approach with respect to this previous attempt is 
twofold. First, we keep all the orders in the $\beta_t$ expansion of the
interaction. Second, and more important, we use the explicit 
knowledge of the first order in the $\beta_s$ expansion to map 
the original 
gauge model to the equivalent Ising model, {\sl explicitly relating the 
gauge coupling and the Ising coupling} in a new original way, 
completely different from that of~\cite{ps}.
 From the knowledge of the 
location of the Ising phase transition we can thus reconstruct the exact 
critical deconfinement temperature.

Let us follow this procedure in the $N_t=1$ case. 
The crucial point of the whole approach is the identification
of the Ising variable 
``embedded'' in the Polyakov loops.
 Let us follow as a  first example the analysis of 
ref.~\cite{ps}. The simplest way to extract these Ising variables is to 
decompose the Polyakov loops:
\eq
P(\vec x)~=~
2 \cos (\theta_{\vec x}) ~,~~~ 
\theta_{\vec x} \in [-\pi,\pi]
~~~.
\en
which are $SU(2)$ variables, into the product of a $Z_2$ variable 
$\sigma(\vec x)$, which is simply a sign, and a SO(3) variable:
\eq
\tilde P(\vec x)~=~
 2 \cos (\theta_{\vec x}) ~,~~~ \theta_{\vec x} \in [-\pi/2,\pi/2]
~~~.
\en
Then by integrating over $\tilde P$ one ends up with the desired 
effective action of the Ising type. All these steps can be easily 
performed if we keep at each stage only the very first order in 
$\beta_t$ (and neglect $\beta_s$). In this case the first step gives us 
the effective action $S_p$ of eq.(\ref{seff}).
Then, it is possible to see that at the first order in $\beta_t$
the SO(3) variables decouple and can be integrated out exactly.
The effective action becomes (neglecting an irrelevant overall constant):
\eq
S_{p}(\beta_t)=\sum_{\vec x}\left\{\beta^{Ising}\sum_{i=1}^d
~\sigma(\vec x)~\sigma(\vec{x}+i)\right\}~~~,
\label{ising1}
\en
with $\beta^{Ising}=\frac{16}{9\pi^2}\beta_t$.
This is exactly the ordinary Ising action which is known to have, both 
in two and three dimensions a second order phase transition located at
$\beta_c^{Ising}=0.44068679\cdots $ and
$\beta_c^{Ising}=0.221652(3)$~\cite{bghp} 
 in two and three  dimensions respectively. From this we obtain the 
following values for the deconfinement temperature:
$\beta_t=\frac{9\pi^2}{16}\beta_c^{Ising}$, namely
$\beta_t=2.447$ and $\beta_t=1.231$ in (2+1) and (3+1) 
dimensions respectively. Even if the order of magnitude is essentially 
correct it is easy to see that these estimates are even worse than the 
plain mean field analysis on the action eq.(\ref{seff}). 
 Notwithstanding this, let us stress again that this approach 
is very interesting in itself, 
because it allows to have a deeper physical insight 
on the mechanism underlying the deconfinement transition.
Let us now improve this analysis by keeping in the various step all the 
orders in the $\beta_t$ expansion and the first order in $\beta_s$.

First, let us start from eq.(\ref{a3}), which is the exact effective 
action, taking into account all orders in $\beta_t$ and let us apply the 
same recipe as above to extract the embedded Ising variables. We obtain 
after some simple algebra:

\eqa
\exp(S_0)&=&  \prod_{\vec x}\left(\prod_{i=1}^d\left\{
\frac{e^{\beta_t \sin(\theta_{\vec x}) \sin(\theta_{\vec x+i})}-
e^{-\beta_t \sin(\theta_{\vec x}) \sin(\theta_{\vec x+i})}}
{2\beta_t \sin(\theta_{\vec x}) \sin(\theta_{\vec x+i})}
\right\}\right)\nonumber\\
&\times&
~e^{\beta_t
\cos(\theta_{\vec x}) \cos(\theta_{\vec x+i})\sigma(\vec x)
\sigma(\vec x+i)}~~.
\label{ising0}
\ena

Let us define 

$$\alpha(\vec x,i)\equiv 
\cos(\theta_{\vec x}) \cos(\theta_{\vec x+i})~~~,$$ 
$$\gamma(\vec x,i)\equiv 
\sin(\theta_{\vec x}) \sin(\theta_{\vec x+i})~~~.$$ 

Since at the deconfinement point the SO(3) variables {\sl are not 
critical}, we can assume that $\theta(\vec x)$ is described also at 
the deconfinement point by a constant master field $\theta_0$ and that 
fluctuations in $\theta$ can be neglected. Let us consistently define:

$$\alpha_0\equiv 
\cos^2(\theta_0)~,~~~~~\gamma_0\equiv 
\sin^2(\theta_0)~~.$$

The problem is thus reduced to the identification of this master field, 
in terms of which we could write $\beta_{t,c}=\beta_c^{Ising}/\alpha_0$

The only way we have to find this master field is to 
identify also at the level of the $O(\beta_s^2)$ effective action
the underlying Ising model.
However 
this is clearly a non-trivial task. In fact
at a first glance one could
think that it is impossible to map our $O(\beta_s^2)$ 
effective action into an equivalent Ising model because the
$O(\beta_s^2)$ contribution is $not$ of the nearest-neighbour type. 
For instance, if we separate also in this case the $Z_2$ degrees of
freedom  from the SO(3) ones as we did above, due to the plaquette structure of
the action, the $Z_2$ degrees of freedom exactly cancel each other, and the
underlying Ising model seems to be lost. 
However there is a completely different way to recognize such an Ising model in
the $O(\beta_s^2)$ interaction. In fact, 
a remarkable and non-trivial consequence of the particular form of the 
$O(\beta_s^2)$ interaction is that it can be exactly reorganized as the 
first term of the strong coupling expansion of an ordinary nearest-neighbour
Ising model, thus allowing to complete the identification. 
In fact, if we keep as 
above a constant value of $\theta=\theta_0$, then
 the relations $C_{1,1}=C_{2,2}$ and 
$C_{1,2}=C_{2,1}$, and above all the fact that $C_{1,1}+C_{1,2}=1$, 
allow us
 to interpret eq.(\ref{adj2}) as the strong coupling expansion of a 
$d$--dimensional Ising model with action, say,
$S_{Ising}=J~\sigma(\vec x)\sigma(\vec x +i)$
 where the values $i=1$ and $i=2$ of the 
indices of $C_{i,j}$ denote the $+1$ and $-1$ values of the Ising 
spin $\sigma(x)$ and the Ising coupling $J$ is given by 
\eq
J=\frac{1}{2}\log\left(\frac{C_{1,1}}{C_{1,2}}\right)=
\frac{1}{2}\log\frac{ 2\beta_t \gamma_0 - 
 \left( 1 - e^{-2\beta_t \gamma_0}
\right)}
{1 - e^{-2\beta_t \gamma_0} 
\left(1+ 2\beta_t \gamma_0\right)}~~~.
\label{e10}
\en
This induced Ising model can be considered as the replica at 
the first order in $\beta_s$ of 
that described above at the zeroth order in $\beta_s$. 
Again we must require  the coupling $J$, to be at its critical value 
$J_c.$
Solving eq.(\ref{e10}) with respect to $\gamma_0$, we find 
$\gamma_0=1.3957/\beta_{t,c}$ in (2+1) dimensions and 
$\gamma_0=0.67383/\beta_{t,c}$ in (3+1) dimensions. Combining these 
values with the above discussed relation: 
$\beta_{t,c}=\beta_c^{Ising}/\alpha_0$, we finally find:

(2+1) dimensions:
$$\beta_t=1.836~,~~~ \alpha_0=0.2398~,~~~\theta_0=0.337\pi~~~$$

(3+1) dimensions:
$$\beta_t=0.8954~,~~~ \alpha_0=0.2475~,~~~\theta_0=0.334\pi~~~.$$

As we anticipated above, these values for 
$\beta_c$ are much lower than those of ref.~\cite{ps}, and the one for 
$d=3$ is in good agreement with the result obtained by Montecarlo simulations.

\section{Conclusions}
The approach outlined in sect.4 can be extended also to 
$N_t>1$~\cite{bcdp}. In 
following this extension one must take care of some non-trivial 
features of the models, like the fact that 
 the critical coupling $\beta_c$ as a function of $N_t$
obeys different scaling laws in (2+1) and (3+1) dimensions.
However the pattern of our approach needs not to be changed. 
The agreement with the  Montecarlo results (when they exist) remains 
very good. Since our approach is not limited by the magnitude of $N_t$,
we can hope that, as $N_t$ increases, a sensible continuum limit
for $T_c$ could be taken. To reach this goal we must
be able to reconstruct the correct scaling laws in the large $N_t$ limit.
This is certainly possible for the (2+1) dimensional model (see~\cite{bcdp} for
details), but it is still an open problem in the (3+1) dimensional case.
In any case, besides the numerical results, we think that the improvements
that we have discussed in this contribution both in constructing the effective
action and in extracting the critical coupling
can help us to have a better and deeper understanding of the physics of the
deconfinement transition in lattice gauge theories.

\vskip 1cm
\centerline{{\bf  Acknowledgements}}
\vskip 0.5cm
We thank M. Bill\'o, A. D'Adda, F. Gliozzi and S.Panzeri 
for many helpful discussions.

%Appendix

\vskip 1cm
\appendix{\Large {\bf{Appendix}}}
\vskip 0.5cm
\renewcommand{\theequation}{A.\arabic{equation}}
\setcounter{equation}{0}

In this Appendix we shall evaluate  the matrix elements 
$C_{kl}$.

To this end let us select in the sum over representations contained in 
$C_{kl}$ the $j^{th}$ term:
\eq
A_{kl}^j\equiv
\int D~U |U^{kl}|^2\chi_j(V(\vec x)~U~V^\dagger(\vec{x}+i)
U^\dagger)
\label{ab8}
\en
so that $C_{kl}$ can be written as:
\eq
C_{kl}=\sum_{r=0}^\infty
d_r~\left[\frac{I_{2r+1}(\beta_t)}{I_1(\beta_t)}\right]^{N_t}~
A^r_{kl}~~~.
\label{ab8bis}
\en

Let us use the following relation:
\eq
\chi_j(U)=\sum_{k=0}^{[j]}\frac{(-1)^k(2j-k)!}{k!~(2j-2k)!}
\chi_{\frac{1}{2}}^{2j-2k}~,~~~j=0,\frac{1}{2},1,\cdots
\label{ab2}
\en
(where $[j]$ denotes the integer part of $j$),
which is a direct consequence of the identification of the $SU(2)$ 
characters with the Tchebichef polynomials of second kind: $\chi_j(U)=
U_{2j}(\cos(\theta))$ (where $\theta$ denotes, as usual, the
 invariant angle of the matrix $U$).

We can rewrite (\ref{ab8}) as:
\eq
A_{kl}^j=\sum_{k=0}^{[j]}\frac{(-1)^k(2j-k)!}{k!~(2j-2k)!}
\int D~U |U^{kl}|^2\chi_{\frac{1}{2}}^{2j-2k}
(V(\vec x)~U~V^\dagger(\vec{x}+i)
U^\dagger)~~.
\label{ab9}
\en

Since the $U$ matrix elements always appear in the form $|U^{kl}|^2$
(with the indices in the fundamental representation) it turns out that
a very useful parametrization is:
\eq
U=a_0{\bf 1}+\sum_{i=1}^3 a_i\sigma_i
\en
where $\sigma_i$  are the Pauli matrices, the $a_i$ are real numbers 
constrained by: $\sum_{i=0}^3 a^2_i=1$. In this parametrization we have
$|U^{12}|^2=a^2_1+a_2^2$ and 
$|U^{11}|^2=|U^{22}|^2=a^2_0+a_3^2$. Setting $a^2_1+a_2^2=x$
we see that the $\chi_{\frac{1}{2}}$ in eq.(\ref{ab9}) becomes
\eq
\chi_{\frac{1}{2}}
(V(\vec x)~U~V^\dagger(\vec{x}+i)
U^\dagger)=g+hx
\en
with $g=2\cos(\theta_{\vec x}-\theta_{\vec x +i})$ and
 $h=-4\sin\theta_{\vec x}~\sin\theta_{\vec x +i}$. The measure $DU$ 
becomes $dx$, with integration limits $0$ and $1$, according to the 
above mentioned constraint on the $a_i$. The $A_{kl}^j$ integrals become:

\eq
A_{11}^j=A_{22}^j=\sum_{k=0}^{[j]}\frac{(-1)^k(2j-k)!}{k!~(2j-2k)!}
\int_0^1 dx (1-x)  (g+hx)^{2j-2k}
\label{ab10}
\en
\eq
A_{12}^j=\sum_{k=0}^{[j]}\frac{(-1)^k(2j-k)!}{k!~(2j-2k)!}
\int_0^1 dx x  (g+hx)^{2j-2k}~~~.
\label{ab10bis}
\en

Before evaluating these integrals, as a preliminary exercise, let us 
calculate the simpler integral in which no contribution coming from the 
spacelike plaquette is present. Let us call it $A_0^j$:
\eq
A_{0}^j=\sum_{k=0}^{[j]}\frac{(-1)^k(2j-k)!}{k!~(2j-2k)!}
\int_0^1 dx  (g+hx)^{2j-2k}
\label{ab11}
\en

If we are able to evaluate the integrals and sum up the series we shall
find an alternative way to go from eq.(\ref{a1}) to eq.(\ref{a2}).

The integrals in the sum of eq.(\ref{ab11}) can be done 
straightforwardly:
\eq
\int_0^1dx (g+hx)^n=\frac{(g+h)^{n+1}-g^{n+1}}{h(n+1)}
\en
inserting this result in eq.(\ref{ab11}), and using the explicit 
expression for $g$ and $h$ we have:
\eq
A_{0}^j=\sum_{k=0}^{[j]}\frac{(-1)^k(2j-k)!}{k!~(2j+1-2k)!}
\frac{[2\cos(\theta_{\vec x}+\theta_{\vec x +i})]^{n+1}-
[2\cos(\theta_{\vec x}-\theta_{\vec x +i})]^{n+1}}
{-4\sin\theta_{\vec x}~\sin\theta_{\vec x +i}}~. 
\label{ab11bis}
\en

By using the explicit expression of the Tchebichef polynomials of first 
type:
\eq
T_{n}(\cos(\theta))=\cos(n\theta)
=\frac{n}{2}\sum_{k=0}^{[n/2]}\frac{(-1)^k(n-k-1)!}{k!~(n-2k)!}
[\cos(n\theta)]^{2j-2k}~,
\label{ab12}
\en
we can rewrite $A_0^j$ as 
\eq
A_{0}^j=
\frac{\cos((2j+1)(\theta_{\vec x}-\theta_{\vec x +i}))-
\cos((2j+1)(\theta_{\vec x}+\theta_{\vec x +i}))}
{2(2j+1)\sin\theta_{\vec x}~\sin\theta_{\vec x +i}} 
\label{ab13}
\en
using eq.(\ref{a2ter}), inserting the result in the sum on the 
representations, and using the explicit expression for the characters 
$\chi_j$ we exactly obtain eq.(\ref{a2}).

Let us now  calculate $A_{11}^j$. Also in this case the integral 
contained in eq.(\ref{ab10}) are straightforward:

\eq
\int_0^1dx (1-x)(g+hx)^n=\frac{(g+h)^{n+2}-g^{n+2}}
{h^2(n+1)(n+2)}-\frac{g^{n+1}}{h(n+1)}~~.
\label{ab15}
\en

The second term in eq.(\ref{ab15}) can be treated exactly as we did 
above for $A_0^j$. The contribution to $A_{11}^j$ coming from it 
is (see eq.(\ref{ab13})):
\eq
\frac{\cos((2j+1)(\theta_{\vec x}-\theta_{\vec x +i}))}
{2(2j+1)\sin\theta_{\vec x}~\sin\theta_{\vec x +i}}~~. 
\label{ab16}
\en

The first term of eq.(\ref{ab15}), after inserting the expression for 
$g$ and $h$ gives:
\eq
\sum_{k=0}^{[j]}\frac{(-1)^k(2j-k)!}{k!~(2j+2-2k)!}
\frac{[2\cos(\theta_{\vec x}+\theta_{\vec x +i})]^{n+2}-
[2\cos(\theta_{\vec x}-\theta_{\vec x +i})]^{n+2}}
{16(\sin\theta_{\vec x})^2(\sin\theta_{\vec x +i})^2}~. 
\label{ab17}
\en
To sum this series the following trick is needed. Let us divide and 
multiply for $2j+1$ and let us split this factor at the numerator as 
$(2j+1-k)+k$. 
Then the sum (\ref{ab17}) is splitted into two sums that, after suitable 
redefinition of the indices can be reduced to the sum (\ref{ab12}). 
Collecting together all the pieces one finds:

\eqa
A_{11}^j&=&
\frac{\cos((2j+2)(\theta_{\vec x}+\theta_{\vec x +i}))-
\cos((2j+2)(\theta_{\vec x}-\theta_{\vec x +i}))}
{8(2j+1)(2j+2)(\sin\theta_{\vec x})^2(\sin\theta_{\vec x +i})^2}
\nonumber \\
&-&\frac{\cos((2j)(\theta_{\vec x}+\theta_{\vec x +i}))-
\cos((2j)(\theta_{\vec x}-\theta_{\vec x +i}))}
{8(2j+1)(2j)(\sin\theta_{\vec x})^2(\sin\theta_{\vec x +i})^2}
\nonumber \\
&+&\frac{\cos((2j+1)(\theta_{\vec x}-\theta_{\vec x +i}))}
{2(2j+1)\sin\theta_{\vec x}~\sin\theta_{\vec x +i}}~~~.
\label{ab18}
\ena

By using the definitions (\ref{ab10}),(\ref{ab10bis}) and (\ref{ab11})
one can immediately obtain $A^j_{12}$ as the difference:
$A^j_{12}=A^j_{0}-A^j_{11}$ .

Inserting these results in eq.(\ref{ab8bis}) one finally 
obtains the $C_{kl}$ coefficients. Simple trigonometric relations allow 
to write these coefficients in a compact form:

\eq
C_{11}=C_{22}=
\frac{1}{2}(C_0+C_1)
\en
\eq
C_{12}=C_{21}=
\frac{1}{2}(C_0-C_1)
\en
with:
\eqa
C_1 & = & {1\over {2\sin\theta_{\vec x}\sin\theta_{\vec x +i}}}
\biggl\{\sum_{j=0}^\infty 
\biggl[\chi_j(\theta_{\vec x})\chi_j(\theta_{\vec x+i})
\frac{I_{2j+2}(\beta_t)^{N_t}-I_{2j}(\beta_t)^{N_t}}
{(2j+1)I_1(\beta_t)^{N_t}}\nonumber\\
& & +2\cos\left[(2j+1)\theta_{\vec x}\right]\cos\left[(2j+1)\theta_{\vec x +i}
\right]
\left(\frac{I_{2j+1}(\beta_t)}{I_1(\beta_t)}\right)^{N_t}\biggr]+
\left[\frac{I_0(\beta_t)}{I_1(\beta_t)}\right]^{N_t}\biggr\}~~.
\ena

\end{document}